\documentclass[manuscript]{aastex}
\usepackage{epsfig}
\usepackage{amsmath}
\begin{document}

\title{The Possible Interstellar Anion CH$_2$CN$^-$: Spectroscopic Constants,
Vibrational Frequencies, and Other Considerations}

\date{\today}
\author{Ryan C. Fortenberry}
\email{Ryan.C.Fortenberry@nasa.gov}
\affil{NASA Ames Research Center, Moffett Field, California
94035-1000, U.S.A.}
\author{T. Daniel Crawford}
\affil{Department of Chemistry, Virginia Tech, Blacksburg,
Virginia 24061, U.S.A.}
\author{Timothy J. Lee}
\email{Timothy.J.Lee@nasa.gov}
\affil{NASA Ames Research Center, Moffett Field, California
94035-1000, U.S.A.}

\begin{abstract}

It is hypothesized that the $A\ ^1B_1 \leftarrow \tilde{X} ^1A'$ excitation
into the dipole-bound state of the cyanomethyl anion (CH$_2$CN$^-$) is propsed
as the carrier for one diffuse interstellar band.  However, this particular
molecular system has not been detected in the interstellar medium even though
the related cyanomethyl radical and the isoelectronic ketenimine molecule have
been found.  In this study we are employing the use of proven quartic force
fields and second-order vibrational perturbation theory to compute accurate
spectroscopic constants and fundamental vibrational frequencies for $^1A'$
CH$_2$CN$^-$ in order to assist in laboratory studies and astronomical
observations.

\end{abstract}

\maketitle

{\bf{Keywords:}} Astrochemistry, ISM: molecular anions, Quartic force fields,
Rotational constants, Vibrational frequencies

\section{Introduction}

It has been recently shown that anions are more prevalent in the interstellar
medium (ISM) than previously thought. C$_n$H$^-$ where $n=4,6,8$
\citep{McCarthy06, Cernicharo07, Brunken07, Remijan07, Cordiner11} and the
isoelectronic C$_n$N$^-$ set where $n=1,3,5$ \citep{Thaddeus08, Cernicharo08,
Agundez10} have been conclusively detected in the circumstellar envelope of the
carbon-rich star IRC+10 216 and in a growing number of other interstellar
regions, as well \citep{Cordiner11}.  It has also been proposed that anions
\citep{Sarre00, Cordiner07} may play a role in explaining some features of the
diffuse interstellar bands (DIBs), the unattributed interstellar spectrum
nearly ubiquitously detected in interstellar sightlines for almost a century
\citep{Sarre06}.  The $\lambda$ 8037 \AA\ DIB, in particular, shows strong
correlation to the dipole-bound excited state transition ($A\ ^1B_1 \leftarrow
\tilde{X}\ ^1A'$) of CH$_2$CN$^-$.  The radical form of this anion
\citep{Irvine88} and the isoelectronic ketenimine (CH$_2$CNH) \citep{Lovas06}
are known to exist in the ISM, and various interstellar processes could lead to
the creation of CH$_2$CN$^-$ \citep{Sailer03, Andreazza06, Cordiner07,
Romanskii08a, Romanskii08b, Yang10, Larsson12}.  However, no conclusive
evidence for the interstellar presence of this anion has yet been reported in
the literature.

Dipole-bound states of anions represent a relatively new field of exploration
for astrochemistry.  An extra electron can be bound to the system by a simple
monopole-dipole interaction in exactly one state if the dipole moment is large
enough.  The minimum dipole value is at least 1.625 D \citep{Fermi47}, but, in
reality, it is probably closer to 2.5 D \citep{Gutsev95}.  For systems like
CH$_2$CN$^-$, the ground state anion is closed-shell, and the additional
electron is bound by valence forces and not dipolar forces.  The resulting
closed-shell anion is relatively stable \citep{Simons08, Lykke87, Gutsev95}.
Additionally, the valence ground state allows for excitation into a single
excited state of dipole-bound character.  Recent studies have suggested other
anions that possess dipole-bound excited states and some that even possess
further excited valence states in addition to dipole-bound ones
\citep{Fortenberry11dbs, Fortenberry11Rev, Fortenberry112dbs,
Fortenberry123dbs}.  More studies have examined CH$_2$CN$^-$ than any other
closed-shell anion known to possess a dipole-bound excited state except for
possibly CH$_2$CHO$^-$ \citep{Wetmore80, Mullin92, Mullin93, Simons08}.  A
detection of one of these anions in the ISM could give new insights into
currently unresolved electronic interstellar spectra. 

Even though the electronic spectrum of CH$_2$CN$^-$ motivates the astronomical
interest in this molecule, the most reliable means of interstellar detection
for a specific molecule remains rotational spectroscopy \citep{McCarthy01}.
Recent computational studies have been able to provide the necessary
spectroscopic constants to aid laboratory microwave studies for the rotational
spectrum of NCO$^-$ \citep{Lattanzi10}.  Other computational results have even
been accurate enough to provide reference data for C$_5$N$^-$ \citep{Aoki00,
Botschwina05} that led to its interstellar detection \citep{Cernicharo08}.
Using established quantum chemical computational tools \citep{Huang08} proven
to provide rotational constants often as accurate as 20 MHz (especially for the
$B$- and $C$-type constants) and also vibrational frequencies accurate to 5
cm$^{-1}$ or better \citep{Huang08, Huang09, Huang11, Huang11D, Inostroza11,
Fortenberry11HOCO, Fortenberry11cHOCO, Fortenberry12hococat,
Fortenberry12hocscat, Fortenberry12HCN}, we are computing the spectroscopic
constants and also the fundamental vibrational frequencies to assist in the
detection of CH$_2$CN$^-$ in its ground $^1A'$ state.  The data provided here
will inform experimental study in the laboratory and astonomical observation in
the ISM of this anion.

\section{Computational Details}

Quartic force fields (QFFs) have been the primary low-cost means by which
accurate computations of spectroscopic constants and vibrational frequencies
have been determined.  QFFs are fourth-order Taylor series approximations to the
anharmonic potential for a given system of interest:
\begin{equation}
V=\frac{1}{2}\sum_{ij}F_{ij}\Delta_i\Delta_j +
\frac{1}{6}\sum_{ijk}F_{ikj}\Delta_i\Delta_j\Delta_k +
\frac{1}{24}\sum_{ijkl}F_{ikjl}\Delta_i\Delta_j\Delta_k\Delta_l
\label{VVib}
\end{equation}
for the force constants, $F_{ij\ldots}$, and displacements, $\Delta_i$.  Unlike
the corresponding radical, CH$_2$CN$^-$ is a non-planar $C_s$ system.  It has
nine vibrational degrees of freedom.  The symmetry-internal coordinates are
defined from the atoms labeled in Fig.~\ref{fig} as:
\begin{align}
S_1(A') &=(\mathrm{C_1}-\mathrm{H_1})+(\mathrm{C_2}-\mathrm{H_2})\\
S_2(A') &=\mathrm{C_1}-\mathrm{C_2}\\
S_3(A') &=\mathrm{C_2}-\mathrm{N}\\
S_4(A') &=(\angle \mathrm{C_2}-\mathrm{C_1}-\mathrm{H_1})+(\angle \mathrm{C_2}-\mathrm{C_1}-\mathrm{H_2})\\
S_5(A') &=\angle \mathrm{H_1}-\mathrm{C_1}-\mathrm{H_2}\\
S_6(A') &=\angle \mathrm{C_1}-\mathrm{C_2}-\mathrm{N}-\mathrm{X}\\
S_7(A'') &=(\mathrm{C_1}-\mathrm{H_1})-(\mathrm{C_2}-\mathrm{H_2})\\
S_8(A'') &=(\angle \mathrm{C_2}-\mathrm{C_1}-\mathrm{H_1})-(\angle \mathrm{C_2}-\mathrm{C_1}-\mathrm{H_2})\\
S_9(A'') &=\angle \mathrm{C_1}-\mathrm{C_2}-\mathrm{N}-\mathrm{X}
\end{align}
%
where X is a dummy atom located in the $\angle$C$_1-$C$_2-$N plane directly
above C$_2$.  This point is necessary to define the in- and out-of-plane
bending in $S_6$ and $S_9$ with linear coordinates LINX and LINY, respectively,
available in the INTDER program \citep{intder}. These coordinates are discussed
in more detail in \cite{Fortenberry12hococat}.  Due to the near-linearity in
$\angle$C$_1-$C$_2-$N, the dihedral angles ($\angle$N$-$C$_2-$C$_1-$H$_{1/2}$)
are not well-defined in this system.  However, linear combinations of
coordinates $S_4-S_9$ account for them in the coordinate system given.

The QFF is determined from points on the potential surface where up to a total
of four displacements for each of the symmetry-internal coordinates are
combined.  The individual displacements are 0.005 \AA\ for bond lengths and
0.005 radians for bond angles and linear bends.  This produces 1814
symmetry-unique points.  At each point, spin-restricted Hartree-Fock (RHF)
\citep{ScheinerRHF87} referenced coupled cluster \citep{Lee95Accu, ccreview,
Shavitt09} singles, doubles, and perturbative triples, CCSD(T) \citep{Rag89},
energies are computed with Dunning's aug-cc-pVTZ and aug-cc-pVQZ basis sets
\citep{aug-cc-pVXZ, Dunning01}.  It has been shown that ground state
computations of anions can be accurately undertaken with only adding a single
set of functions to account for orbital diffuseness \citep{Lee97, Lee99,
Skurski00}.  The triple- and quadruple-zeta energies at each point are then
extrapolated to the complete one-particle basis set (CBS) limit using a
two-point formula \citep{Helgaker97}.  To the extrapolated energies, further
corrections are added for core-correlation \citep{Martin94} and scalar
relativistic effects \citep{Douglas74}.  The use of these three terms in the
composite energy gives the CcCR QFF, which is so named for CBS extrapolation
(C), core-correlation (cC), and relativistic effects (R) included.  This QFF
differs from the CcCR QFF utilized effectively in previous studies
\citep{Huang08, Huang09, Huang11, Inostroza11, Fortenberry11HOCO,
Fortenberry11cHOCO, Fortenberry12hococat, Fortenberry12hocscat,
Fortenberry12HCN} only in that aug-cc-pV5Z energies are not included in the
basis set extrapolation.

The CcCR energies from the 2339 total symmetry-redundant points fit to a sum of
residual squares on the order of 10$^{-16}$ a.u.$^2$ yield the QFF.  The
symmetry-internal coordinate QFF is transformed into Cartesian coordinates with
the INTDER program \citep{intder}.  The spectroscopic constants and vibrational
frequencies are computed with second-order vibrational perturbation theory
(VPT) \citep{Mills72, Watson77, Papousek82} from a modified version of the
SPECTRO \citep{spectro} program.

\section{Discussion}

The vibrationally-averaged $R_{\alpha}$ structure of the cyanomethyl anion
(CH$_2$CN$^-$) is shown in Fig.~\ref{fig}.  The $C_s$ anion geometry only
differs slightly from the $C_{2v}$ radical.  The singly-occupied $p$ orbital
above and below the $sp^2$ methyl carbon in the radical allows for a planar
$C_{2v}$ molecule, but the addition of the extra electron to create a
closed-shell system leads to interference from the other already occupied
orbitals.  As a result, the bonds involving the methyl group bend the molecule
into an $sp^3$ pyramidal structure reminiscent of ammonia with the extra
lone-pair occupying the previously singly-occupied $p$ orbital.  The off-axis
dihedrals showcase this effect as they are $\pm 109.726^{\circ}$.  The
dihedrals and the other geometrical parameters are listed in Table
\ref{StructHarm} for both the vibrationally-averaged (zero-point) and
equilibrium structures.

The fundamental vibrational frequencies of CH$_2$CN$^-$, given in Table
\ref{vptvci}, appear to be well-behaved and provide reference data for infrared
studies.  The lone exception is the $\nu_9$ C$-$C$-$N out-of-plane bend.  When
progressing from the harmonic approximation to the anharmonic, the frequency is
reduced by 54.2\% from 342.2 cm$^{-1}$ to 157.1 cm$^{-1}$.  It is known that
there is issue with VPT when describing modes of strong anharmonicity
\citep{Dateo94, Martin99, Torrent05, Yurchenko09}.  Hence, the anharmonic VPT
$\nu_9$ frequency is probably not as accurate as the other anharmonic
fundamentals provided.  In addition to the fundamentals, VPT also produces the
zero-point and equilibrium spectroscopic constants for the system of study.
These and most of the other spectroscopic constants for CH$_2$CN$^-$ are given
in Table \ref{StructHarm} with the vibration-rotation interaction constants
listed in Table \ref{vib-rot}.  Additionally, the symmetry-internal coordinate
force constants computed are given in Tables \ref{fc1} and \ref{fc2}.

The possible interstellar microwave detection of CH$_2$CN$^-$ could be
referenced by related and potentially more abundant species.  As mentioned
previously, the ketenimine (CH$_2$CNH) molecule, which is isoelectronic to the
cyanomethyl anion, has been detected in the ISM.  The extra out-of-plane
hydrogen in ketenimine places the $A_0$ (201 445.279 MHz), $B_0$ (9 663.159 3
MHz), and $C_0$ (9 470.154 7 MHz) experimental rotational constants
\citep{Bane11} below those of CH$_2$CN$^-$ by 32 500 MHz, 1 160 MHz, and 939
MHz, respectively.  The rotational constants discussed for CH$_2$CN$^-$ are the
vibrationally-averaged values.  Additionally, the known interstellar molecule
propadienylidene (CH$_2$CC) \citep{Cernicharo91} has been proposed as a
possible DIB carrier by \cite{Maier11}, and its geometry is very similar in
structure to CH$_2$CN.  The CH$_2$CC $A$-, $B$-, and $C$-type rotational
constants are 288 783 MHz, 10 588.639 MHz, and 10 203.966 MHz, respectively,
\citep{Vrtilek90}, but the pyramidalization of the methyl carbon in the anion
of interest consequently gives a smaller $A$-type rotational constant and larger 
$B$- and $C-$type constants relative to CH$_2$CC.  The differences in the
rotational spectra of ketenimine and propadienylidene may help to serve as
markers for the microwave detection of CH$_2$CN$^-$ in the ISM.

The structures of CH$_2$CC and the CH$_2$CN radical are very similar, and the
relationship between their rotational constants and those of CH$_2$CN$^-$ is
also similar.  For the CH$_2$CN radical, the $A_0$ rotational constant is 284
981 MHz, $B_0$ is 10 426.765 MHz, and $C_0$ is 9 876.035 MHz \citep{Saito97}.
Previously computed $A$-type CcCR rotational constants for other quasi-linear
systems have not been as close to experimental values as were the other two
rotational constants \citep{Inostroza11, Fortenberry11HOCO, Fortenberry11cHOCO,
Fortenberry12hococat, Fortenberry12hocscat}, but the $A_0$ constant computed
here should still be accurate to 500 MHz or better.  This should be accurate
enough to make comparison between the cyanomethyl radical and anion.  From our
results, we can conclude that $A_0$ for CH$_2$CN$^-$ should be as much as 50
GHz less than that of the radical.  Conversely, the anion $B_0$ and $C_0$
constants, where accuracies are more than an order of magnitude better, should
be 400 and 500 MHz, respectively, greater than those of the radical.  Later
work by \cite{Ozeki04} indicates that $B-C$ for the radical should be 370.735 9
MHz whereas the earlier estimate by \cite{Saito97} is 550.730 MHz.  Our
computations give $B-C$ of CH$_2$CN$^-$ to be 437.21 MHz.  Hence $B-C$ is not a
good means by which one should distinguish the radical from the anion.  The
microwave spectra of both the cyanomethyl radical and anion are closely
related, but our data should allow the two to be distinguished in observation.
The dipole moment for the anion (0.90 D; see Table \ref{StructHarm}) indicates
that a rotational spectrum should possess adequate intensities for detection
provided the interstellar abundance is large enough.  Additionally, the other
spectroscopic constants provided here should aid in further studies of
CH$_2$CN$^-$.

As a final consideration, it is possible that some part of the CH$_2$CN
rotational signal observed could be the result of the $A\ ^1B_1$ dipole-bound
excited state of CH$_2$CN$^-$.  By its definition, the dipole-bound state of an
anion is structurally identical to the neutral since the extra, excited
electron is nearly non-interacting with the rest of the system
\citep{Simons08}.  This behavior has been computationally documented for
various anions with dipole-bound states including CH$_2$CN$^-$
\citep{Fortenberry11dbs, Fortenberry11Rev, Fortenberry112dbs,
Fortenberry123dbs}.  As a result, the rotational spectra of $\tilde{X}\ ^2B_1$
CH$_2$CN and $A\ ^1B_1$ CH$_2$CN$^-$ should be nearly indistinguishable.  This
would obscure a clear marker of the $A\ ^1B_1 \leftarrow \tilde{X} ^1A'$
transition of CH$_2$CN$^-$ as the carrier of the $\lambda$ 8037 \AA\ DIB since
the excited state could not be detected through microwave spectroscopy.  The
anion's dipole-bound excited state vibrational progressions would also be
nearly identical to that of the ground state radical, as well, since
perturbations to the electronic potential would be little affected by the extra
electron in such a highly diffuse orbital.  However, the rotational lines of
$\tilde{X} ^1A'$ CH$_2$CN$^-$ are distinguishable from the radical and would
allow this anion to be detected at least in its ground electronic state.



\section{Acknowledgements}

\begin{figure}
\caption{Top-down and side-on views for the CcCR equilibrium geomtrey of
$\tilde{X}\ ^1A'$ CH$_2$CN$^-$.  $\angle$C$-$C$-$N is shown explicity (units are
degress) in the lower side-on view.}
\includegraphics[width = 6.0 in]{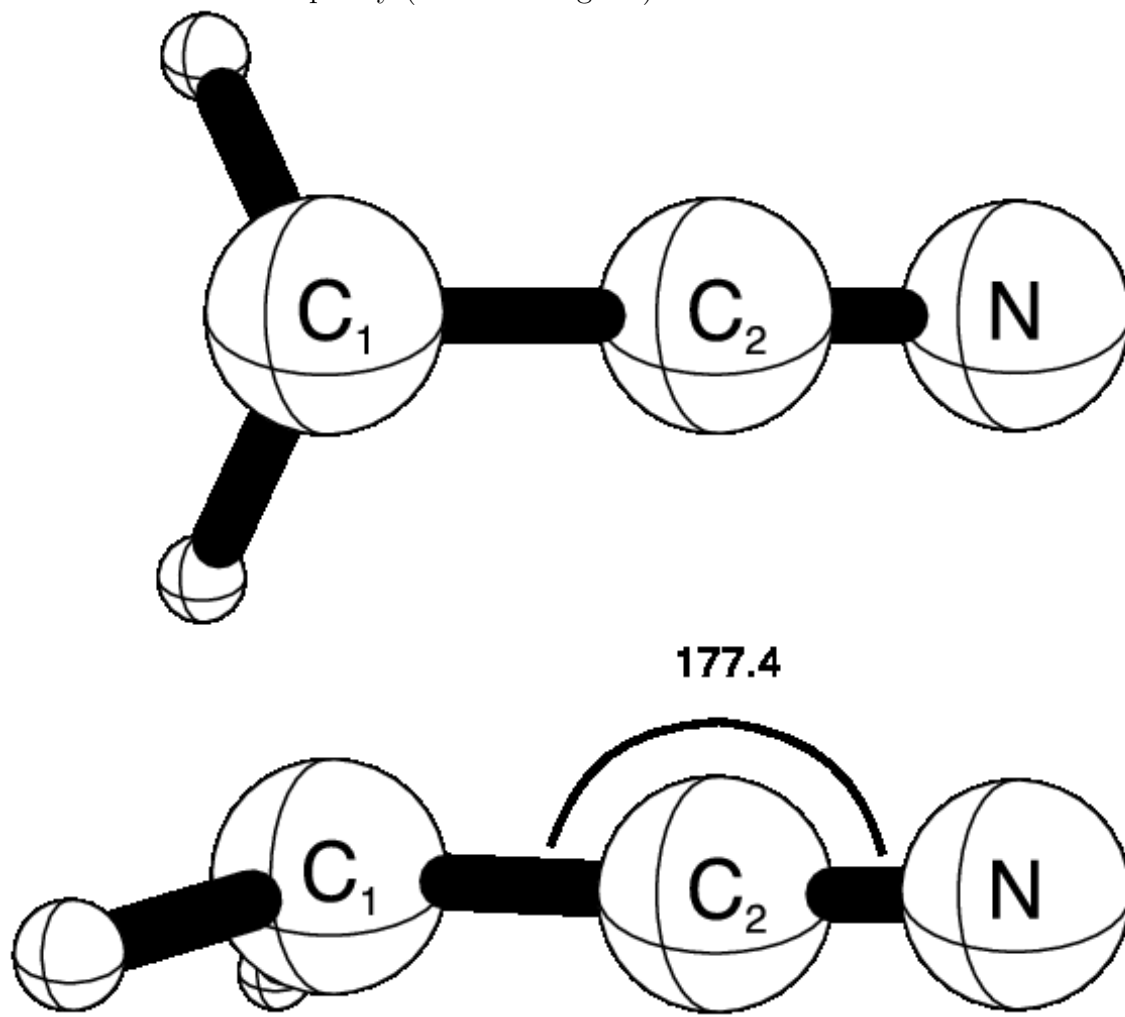}
\label{fig}
\end{figure}

RCF was funded, in part, through the NASA Postdoctoral Program administered by
Oak Ridge Associated Universities.  The computational hardware utilized in this
work was made available by TDC from the U.S. National Science Foundation (NSF)
Multi-User Chemistry Research Instrumentation and Facility (CRIF:MU) award
CHE-0741927.  TDC also acknowledges funding from NSF grant CHE-1058420.  The
work undertaken by TJL was made possible through NASA Grant 10-APRA10-0167.
The CheMVP program was used to created Fig.~\ref{fig}.  The authors thank
Dr.~Julia Rice of IBM Almaden for assistance with editing the manuscript.  RCF
would also like to acknowledge Dr.~Martin Cordiner of the NASA Goddard Space
Flight Center for initiating the discussions that led to the execution of this
project.

\bibliographystyle{apj}
\bibliography{refs}

\begin{thebibliography}{72}
\expandafter\ifx\csname natexlab\endcsname\relax\def\natexlab#1{#1}\fi

\bibitem[{Ag\`undez {et~al.}(2010)Ag\`undez, Cernicharo, Gu\`elin, Kahane,
  Roueff, Klos, Aoiz, Lique, Marcelino, Goicoechea, {Gonz\`alez Garcia},
  Gottlieb, McCarthy, \& Thaddeus}]{Agundez10}
Ag\`undez, M., Cernicharo, J., Gu\`elin, M., {et~al.} 2010, Astron. Astrophys.,
  517, L2

\bibitem[{Allen \& coworkers(2005)}]{intder}
Allen, W.~D., \& coworkers. 2005, $INTDER\ 2005$ is a general program written
  by W. D. Allen and coworkers, which performs vibrational analysis and
  higher-order non-linear transformations.

\bibitem[{Andreazza {et~al.}(2006)Andreazza, Fitzgerald, \&
  Bowie}]{Andreazza06}
Andreazza, H.~J., Fitzgerald, M., \& Bowie, J.~H. 2006, Org. Biomol. Chem., 4,
  2466

\bibitem[{Aoki(2000)}]{Aoki00}
Aoki, K. 2000, Chem. Phys. Lett., 323, 55

\bibitem[{Bane {et~al.}(2011)Bane, Robertson, Thompson, Appadoo, \&
  McNaughton}]{Bane11}
Bane, M.~K., Robertson, E.~G., Thompson, C.~D., Appadoo, D. R.~T., \&
  McNaughton, D. 2011, J. Chem. Phys., 135, 224306

\bibitem[{Botschwina(2005)}]{Botschwina05}
Botschwina, P. 2005, Mol. Phys., 103, 1441

\bibitem[{Br\"unken {et~al.}(2007)Br\"unken, Gupta, Gottlieb, McCarthy, \&
  Thaddeus}]{Brunken07}
Br\"unken, S., Gupta, H., Gottlieb, C.~A., McCarthy, M.~C., \& Thaddeus, P.
  2007, Astrophys. J., 664, L43

\bibitem[{Cernicharo {et~al.}(1991)Cernicharo, Gottlieb, Gu\'{e}lin, Killian,
  Paubert, Thaddeus, \& Vrtilek}]{Cernicharo91}
Cernicharo, J., Gottlieb, C.~A., Gu\'{e}lin, M., {et~al.} 1991, Astrophys. J.,
  368, L39

\bibitem[{Cernicharo {et~al.}(2007)Cernicharo, Gu\`elin, Ag\`undez, Kawaguchi,
  McCarthy, \& Thaddeus}]{Cernicharo07}
Cernicharo, J., Gu\`elin, M., Ag\`undez, M., {et~al.} 2007, Astron. Astrophys.,
  467, L37

\bibitem[{Cernicharo {et~al.}(2008)Cernicharo, Gu\`elin, Agundez, McCarthy, \&
  Thaddeus}]{Cernicharo08}
Cernicharo, J., Gu\`elin, M., Agundez, M., McCarthy, M.~C., \& Thaddeus, P.
  2008, Astrophys. J., 688, L83

\bibitem[{Cordiner {et~al.}(2011)Cordiner, Charnley, Buckle, Walsh, \&
  Millar}]{Cordiner11}
Cordiner, M.~A., Charnley, S.~B., Buckle, J.~V., Walsh, C., \& Millar, T.~J.
  2011, Astrophys. J. Lett., 730, L18

\bibitem[{Cordiner \& Sarre(2007)}]{Cordiner07}
Cordiner, M.~A., \& Sarre, P.~J. 2007, Astron. Astrophys., 472, 537

\bibitem[{Crawford \& Schaefer(2000)}]{ccreview}
Crawford, T.~D., \& Schaefer, H.~F. 2000, in Reviews in Computational
  Chemistry, ed. K.~B. Lipkowitz \& D.~B. Boyd, Vol.~14 (New York: Wiley),
  33--136

\bibitem[{Dateo {et~al.}(1994)Dateo, Lee, \& Schwenke}]{Dateo94}
Dateo, C.~E., Lee, T.~J., \& Schwenke, D.~W. 1994, J. Chem. Phys., 101, 5853

\bibitem[{Douglas \& Kroll(1974)}]{Douglas74}
Douglas, M., \& Kroll, N. 1974, Ann. Phys., 82, 89

\bibitem[{Dunning {et~al.}(2001)Dunning, Peterson, \& Wilson}]{Dunning01}
Dunning, T.~H., Peterson, K.~A., \& Wilson, A.~K. 2001, J. Chem. Phys., 114,
  9244

\bibitem[{Fermi \& Teller(1947)}]{Fermi47}
Fermi, E., \& Teller, E. 1947, Phys. Rev., 72, 399

\bibitem[{Fortenberry \& Crawford(2011{\natexlab{a}})}]{Fortenberry11Rev}
Fortenberry, R.~C., \& Crawford, T.~D. 2011{\natexlab{a}}, Annu. Rep. Comput.
  Chem., 7, 195

\bibitem[{Fortenberry \& Crawford(2011{\natexlab{b}})}]{Fortenberry112dbs}
---. 2011{\natexlab{b}}, J. Phys. Chem. A, 115, 8119

\bibitem[{Fortenberry \& Crawford(2011{\natexlab{c}})}]{Fortenberry11dbs}
---. 2011{\natexlab{c}}, J. Chem. Phys., 134, 154304

\bibitem[{Fortenberry \& Crawford(2012)}]{Fortenberry123dbs}
---. 2012, $to\ be\ submitted$

\bibitem[{Fortenberry {et~al.}(2012{\natexlab{a}})Fortenberry, Huang, Crawford,
  \& Lee}]{Fortenberry12HCN}
Fortenberry, R.~C., Huang, X., Crawford, T.~D., \& Lee, T.~J.
  2012{\natexlab{a}}, J. Phys. Chem. A., $submitted$

\bibitem[{Fortenberry {et~al.}(2011{\natexlab{a}})Fortenberry, Huang,
  Francisco, Crawford, \& Lee}]{Fortenberry11HOCO}
Fortenberry, R.~C., Huang, X., Francisco, J.~S., Crawford, T.~D., \& Lee, T.~J.
  2011{\natexlab{a}}, J. Chem. Phys., 135, 134301

\bibitem[{Fortenberry {et~al.}(2011{\natexlab{b}})Fortenberry, Huang,
  Francisco, Crawford, \& Lee}]{Fortenberry11cHOCO}
---. 2011{\natexlab{b}}, J. Chem. Phys., 135, 214303

\bibitem[{Fortenberry {et~al.}(2012{\natexlab{b}})Fortenberry, Huang,
  Francisco, Crawford, \& Lee}]{Fortenberry12hocscat}
---. 2012{\natexlab{b}}, J. Phys. Chem. A., 116, 9582

\bibitem[{Fortenberry {et~al.}(2012{\natexlab{c}})Fortenberry, Huang,
  Francisco, Crawford, \& Lee}]{Fortenberry12hococat}
---. 2012{\natexlab{c}}, J. Chem. Phys., 136, 234309

\bibitem[{Gaw {et~al.}(1996)Gaw, Willets, Green, \& Handy}]{spectro}
Gaw, J.~F., Willets, A., Green, W.~H., \& Handy, N.~C. 1996, $SPECTRO\
  program$, version 3.0

\bibitem[{Gutsev \& Adamowicz(1995)}]{Gutsev95}
Gutsev, G., \& Adamowicz, A. 1995, Chem. Phys. Lett., 246, 245

\bibitem[{Helgaker {et~al.}(1997)Helgaker, Klopper, Koch, \& Noga}]{Helgaker97}
Helgaker, T., Klopper, W., Koch, H., \& Noga, J. 1997, J. Chem. Phys., 106,
  9639

\bibitem[{Huang \& Lee(2008)}]{Huang08}
Huang, X., \& Lee, T.~J. 2008, J. Chem. Phys., 129, 044312

\bibitem[{Huang \& Lee(2009)}]{Huang09}
---. 2009, J. Chem. Phys., 131, 104301

\bibitem[{Huang \& Lee(2011)}]{Huang11D}
---. 2011, Astrophys. J., 736, 33

\bibitem[{Huang {et~al.}(2011)Huang, Taylor, \& Lee}]{Huang11}
Huang, X., Taylor, P.~R., \& Lee, T.~J. 2011, J. Phys. Chem. A, 115, 5005

\bibitem[{Inostroza {et~al.}(2011)Inostroza, Huang, \& Lee}]{Inostroza11}
Inostroza, N., Huang, X., \& Lee, T.~J. 2011, J. Chem. Phys., 135, 244310

\bibitem[{Irvine {et~al.}(1988)Irvine, Friberg, Hjalmarson, Ishikawa, Kaifu,
  Kawaguchi, Madden, Matthews, Ohishi, Saito, Suzuki, Thaddeus, Turner,
  Yamamoto, \& Ziurys}]{Irvine88}
Irvine, W.~M., Friberg, P., Hjalmarson, A., {et~al.} 1988, Astrophys. J., 335,
  L89

\bibitem[{Kendall {et~al.}(1992)Kendall, Dunning, \& Harrison}]{aug-cc-pVXZ}
Kendall, R.~A., Dunning, T.~H., \& Harrison, R.~J. 1992, J. Chem. Phys., 96,
  6796

\bibitem[{Larsson {et~al.}(2012)Larsson, Geppert, \& Nyman}]{Larsson12}
Larsson, M., Geppert, W.~D., \& Nyman, G. 2012, Rep. Prog. Phys., 75, 066901

\bibitem[{Lattanzi {et~al.}(2010)Lattanzi, Gottlieb, Thaddeus, Thorwirth, \&
  McCarthy}]{Lattanzi10}
Lattanzi, V., Gottlieb, C.~A., Thaddeus, P., Thorwirth, S., \& McCarthy, M.~C.
  2010, Astrophys. J., 720, 1717

\bibitem[{Lee \& Dateo(1997)}]{Lee97}
Lee, T.~J., \& Dateo, C.~E. 1997, J. Chem. Phys., 107, 10373

\bibitem[{Lee \& Dateo(1999)}]{Lee99}
---. 1999, Spectrochim. Acta, Part A., 55, 739

\bibitem[{Lee \& Scuseria(1995)}]{Lee95Accu}
Lee, T.~J., \& Scuseria, G.~E. 1995, in Quantum Mechanical Electronic Structure
  Calculations with Chemical Accuracy, ed. S.~R. Langhoff (Dordrecht: Kluwer
  Academic Publishers), 47--108

\bibitem[{Lovas {et~al.}(2006)Lovas, Hollis, Remijan, \& Jewell}]{Lovas06}
Lovas, F.~J., Hollis, J.~M., Remijan, A.~J., \& Jewell, P.~R. 2006, Astrophys.
  J., 645, L137

\bibitem[{Lykke {et~al.}(1987)Lykke, Neumark, Andersen, Trapa, \&
  Lineberger}]{Lykke87}
Lykke, K.~R., Neumark, D.~M., Andersen, T., Trapa, V.~J., \& Lineberger, W.~C.
  1987, J. Chem. Phys., 87, 6842

\bibitem[{Maier {et~al.}(2011)Maier, Walker, Bohlender, Mazzotti, Raghunandan,
  Fulara, Garkusha, \& Nagy}]{Maier11}
Maier, J.~P., Walker, G. A.~H., Bohlender, D.~A., {et~al.} 2011, Astrophys. J.,
  726, 41

\bibitem[{Martin \& Taylor(1994)}]{Martin94}
Martin, J. M.~L., \& Taylor, P.~R. 1994, Chem. Phys. Lett., 225, 473

\bibitem[{Martin \& Taylor(1999)}]{Martin99}
---. 1999, Mol. Phys., 96, 681

\bibitem[{McCarthy {et~al.}(2006)McCarthy, Gottlieb, Gupta, \&
  Thaddeus}]{McCarthy06}
McCarthy, M.~C., Gottlieb, C.~A., Gupta, H., \& Thaddeus, P. 2006, Astrophys.
  J., 652, L141

\bibitem[{McCarthy \& Thaddeus(2001)}]{McCarthy01}
McCarthy, M.~C., \& Thaddeus, P. 2001, Chem. Soc. Rev., 30, 177

\bibitem[{Mills(1972)}]{Mills72}
Mills, I.~M. 1972, in Molecular Spectroscopy - Modern Research, ed. K.~N. Rao
  \& C.~W. Mathews (New York: Academic Press)

\bibitem[{Mullin {et~al.}(1993)Mullin, Murray, Schulz, \&
  Lineberger}]{Mullin93}
Mullin, A.~S., Murray, K.~K., Schulz, C.~P., \& Lineberger, W.~C. 1993, J.
  Phys. Chem., 97, 10281

\bibitem[{Mullin {et~al.}(1992)Mullin, Murray, Schulz, Szaflarski, \&
  Lineberger}]{Mullin92}
Mullin, A.~S., Murray, K.~K., Schulz, C.~P., Szaflarski, D.~M., \& Lineberger,
  W.~C. 1992, Chem. Phys., 166, 207

\bibitem[{Ozeki {et~al.}(2004)Ozeki, Hirao, Saito, \& Yamamoto}]{Ozeki04}
Ozeki, H., Hirao, T., Saito, S., \& Yamamoto, S. 2004, Astrophys. J., 617, 680

\bibitem[{Papousek \& Aliev(1982)}]{Papousek82}
Papousek, D., \& Aliev, M.~R. 1982, Molecular Vibration-Rotation Spectra
  (Amsterdam: Elsevier)

\bibitem[{Raghavachari {et~al.}(1989)Raghavachari, Trucks., Pople, \&
  Head-Gordon}]{Rag89}
Raghavachari, K., Trucks., G.~W., Pople, J.~A., \& Head-Gordon, M. 1989, Chem.
  Phys. Lett., 157, 479

\bibitem[{Remijan {et~al.}(2007)Remijan, Hollis, Lovas, Cordiner, Millar,
  Markwick-Kemper, \& Jewell}]{Remijan07}
Remijan, A.~J., Hollis, J.~M., Lovas, F.~J., {et~al.} 2007, Astrophys. J., 664,
  L47

\bibitem[{Romanskii(2008{\natexlab{a}})}]{Romanskii08a}
Romanskii, I.~A. 2008{\natexlab{a}}, Russian Chem. Bull. Int. Ed., 57, 1842

\bibitem[{Romanskii(2008{\natexlab{b}})}]{Romanskii08b}
---. 2008{\natexlab{b}}, Russian Chem. Bull. Int. Ed., 57, 1850

\bibitem[{Sailer {et~al.}(2003)Sailer, Pelc, {Lim\~{a}o-Vieira}, Mason,
  Limtrakul, Scheie, Probst, \& M\"{a}rk}]{Sailer03}
Sailer, W., Pelc, A., {Lim\~{a}o-Vieira}, P., {et~al.} 2003, Chem. Phys. Lett.,
  381, 216

\bibitem[{Saito \& Yamamoto(1997)}]{Saito97}
Saito, S., \& Yamamoto, S. 1997, J. Chem. Phys., 107, 1732

\bibitem[{Sarre(2000)}]{Sarre00}
Sarre, P.~J. 2000, Mon. Not. R. Astron. Soc., 313, L14

\bibitem[{Sarre(2006)}]{Sarre06}
---. 2006, J. Mol. Spectrosc., 238, 1

\bibitem[{Scheiner {et~al.}(1987)Scheiner, Scuseria, Rice, Lee, \& {Schaefer
  III}}]{ScheinerRHF87}
Scheiner, A.~C., Scuseria, G.~E., Rice, J.~E., Lee, T.~J., \& {Schaefer III},
  H.~F. 1987, J. Chem. Phys., 87, 5361

\bibitem[{Shavitt \& Bartlett(2009)}]{Shavitt09}
Shavitt, I., \& Bartlett, R.~J. 2009, Many-Body Methods in Chemistry and
  Physics: MBPT and Coupled-Cluster Theory (Cambridge: Cambridge University
  Press)

\bibitem[{Simons(2008)}]{Simons08}
Simons, J. 2008, J. Phys. Chem. A., 112, 6401

\bibitem[{Skurski {et~al.}(2000)Skurski, Gutowski, \& Simons}]{Skurski00}
Skurski, P., Gutowski, M., \& Simons, J. 2000, Int. Journ. Quantum Chem., 80,
  1024

\bibitem[{Thaddeus {et~al.}(2008)Thaddeus, Gottlieb, Gupta, Br\"{u}nken,
  McCarthy, Ag\`{u}ndez, Gu\`{e}lin, \& Cernicharo}]{Thaddeus08}
Thaddeus, P., Gottlieb, C.~A., Gupta, H., {et~al.} 2008, Astrophys. J., 677,
  1132

\bibitem[{{Torrent-Sucarrat} {et~al.}(2005){Torrent-Sucarrat}, Luis, \&
  Kirtman}]{Torrent05}
{Torrent-Sucarrat}, M., Luis, J.~M., \& Kirtman, B. 2005, J. Chem. Phys., 122,
  204108

\bibitem[{Vrtilek {et~al.}(1990)Vrtilek, Gottlieb, Gottlieb, Killian, \&
  Thaddeus}]{Vrtilek90}
Vrtilek, J.~M., Gottlieb, C.~A., Gottlieb, E.~W., Killian, T.~C., \& Thaddeus,
  P. 1990, Astrophys. J., 364, L53

\bibitem[{Watson(1977)}]{Watson77}
Watson, J. K.~G. 1977, in Vibrational Spectra and Structure, ed. J.~R. During
  (Amsterdam: Elsevier)

\bibitem[{Wetmore {et~al.}(1980)Wetmore, {Schaefer III}, Hiberty, \&
  Brauman}]{Wetmore80}
Wetmore, R.~W., {Schaefer III}, H.~F., Hiberty, P.~C., \& Brauman, J.~I. 1980,
  J. Am. Chem. Soc., 102, 5470

\bibitem[{Yang {et~al.}(2010)Yang, Eichelberger, Carpenter, {Martinez, Jr.},
  Snow, \& Bierbaum}]{Yang10}
Yang, Z., Eichelberger, B., Carpenter, M.~Y., {et~al.} 2010, Astrophys. J.,
  723, 1325

\bibitem[{Yurchenko {et~al.}(2009)Yurchenko, Yachmenev, Thiel, Baum, Giesen,
  Melnikov, \& Jensen}]{Yurchenko09}
Yurchenko, S.~N., Yachmenev, A., Thiel, W., {et~al.} 2009, J. Molec.
  Spectrosc., 257, 57

\end{thebibliography}

\newpage

\renewcommand{\baselinestretch}{1}
\begingroup
\begin{table}[h]

\caption{The Zero-Point ($R_{\alpha}$ vibrationally-averaged) and Equilibrium
Structures, Rotational Constants, CCSD(T)/aug-cc-pVQZ Dipole Moment, and Other
Spectroscopic Constants (including Watson $S$-Reduction terms) of CH$_2$CN$^-$
computed from the CcCR QFF.}

\label{StructHarm}

\centering

\scriptsize

\begin{tabular}{l r}
\hline\hline

r$_0$(C$-$N) & 1.184 912 \AA \\
r$_0$(C$-$C) & 1.394 316 \AA \\
r$_0$(C$-$H) & 1.089 964 \AA \\
$\angle _0($H$-$C$-$C) & 118.378$^\circ$ \\
$\angle _0($C$-$C$-$N) & 177.714$^\circ$ \\
$\angle _0($H$-$C$-$H) & 111.828$^\circ$ \\
$\angle _0$N$-$C$-$C$-$H & $\pm$ 109.726$^\circ$ \\
$A_0$ & 233 945.4 MHz\\
$B_0$ & 10 823.22 MHz\\
$C_0$ & 10 386.01 MHz\\
r$_e$(C$-$N) & 1.186 272 \AA \\
r$_e$(C$-$C) & 1.391 438 \AA \\
r$_e$(C$-$H) & 1.080 053 \AA \\
$\angle _e($H$-$C$-$C) & 116.976$^\circ$ \\
$\angle _e($C$-$C$-$N) & 177.423$^\circ$ \\
$\angle _e($H$-$C$-$H) & 114.048$^\circ$ \\
$\angle _e$H$-$C$-$C$-$N & $\pm$ 109.726$^\circ$ \\
$A_e$ & 230 904.9 MHz\\
$B_e$ & 10 849.79 MHz\\
$C_e$ & 10 445.37 MHz\\
$\mu$ & 0.90 D\\
$\mu_x$$^a$ & 0.86 D\\
$\mu_y$$^a$ & 0.24 D\\
$\tau'_{aaaa}$ & -100.708 MHz \\
$\tau'_{bbbb}$ &   -0.020 MHz \\
$\tau'_{cccc}$ &   -0.017 MHz \\
$\tau'_{aabb}$ &   -1.812 MHz \\
$\tau'_{aacc}$ &    0.043 MHz \\
$\tau'_{bbcc}$ &   -0.018 MHz \\
$\Phi_{aaa}$ & 1.397 $\times 10^4$ Hz \\ 
$\Phi_{bbb}$ & 0.000 Hz \\
$\Phi_{ccc}$ & 0.001 Hz \\
$\Phi_{aab}$ & 21.402 Hz \\
$\Phi_{abb}$ & 3.346 Hz \\
$\Phi_{aac}$ & -296.063 Hz \\
$\Phi_{bbc}$ & -0.002 Hz \\
$\Phi_{acc}$ & -0.016 Hz \\
$\Phi_{bcc}$ &  0.001 Hz \\
$\Phi_{abc}$ &  1.224 Hz \\
$D_J$   &  0.005 MHz \\
$D_{JK}$&  0.434 MHz \\
$D_K$   & 24.739 MHz \\
$d_1$   &  0.000 MHz \\
$d_2$   &  0.000 MHz \\
$H_J$    & -0.001 Hz \\
$H_{JK}$ &  2.297 Hz\\
$H_{KJ}$ &  -278.570 Hz\\
$H_K$    &  1.424 $\times 10^4$ Hz\\
$h_1$    &  -0.001 Hz\\
$h_2$    &  0.001 Hz\\
$h_3$    &  0.000 Hz\\

\hline
\end{tabular}

$^a$The CH$_2$CN$^-$ coordinates (with the center-of-mass at the origin)
utilized to generated the dipole moment components are: C$_1$, 1.25313,
0.058510, 0.000000; C$_2$ -0.136782, -0.005295, 0.000000; N, -1.216834,
-0.006267, 0.000000; H, 1.807100, -0.273271, $\pm$0.995163.

\end{table}
\endgroup
\renewcommand{\baselinestretch}{2}

\begingroup
\begin{table}[h]

\caption{The VPT CcCR QFF fundamental vibrational frequencies (in cm$^{-1}$)
for CH$_2$CN$^-$.} 

\centering
\begin{tabular}{c l | c c}
\hline\hline
\label{vptvci}

Mode & \multicolumn{1}{c|}{Description} & \hspace{0.1in}Harmonic\hspace{0.1in}
& \hspace{0.1in}Anharmonic\hspace{0.1in} \\

\hline

$\nu_1(A')$ & symmetric C$-$H stretch  & 3122.3 & 2987.0\\
$\nu_2(A')$ & C$-$N stretch            & 2115.7 & 2100.3\\
$\nu_3(A')$ & H$-$C$-$H symmetric bend & 1298.2 & 1262.2\\
$\nu_4(A')$ & C$-$C stretch            & 991.5 & 956.1\\
$\nu_5(A')$ & C$-$C$-$N in-plane bend  & 585.5 & 556.9\\
$\nu_6(A')$ & symmetric torsion        & 434.5 & 433.1\\
$\nu_7(A'')$ & antisymmetric C$-$H stretch & 3191.3 & 3045.1\\
$\nu_8(A'')$ & antisymmetric torsion       & 1045.5 & 1033.8\\
$\nu_9(A'')$ & C$-$C$-$N out-of-plane bend & 343.2 & 157.1\\

\hline\hline
\end{tabular}

\end{table}
\endgroup

\begingroup
\begin{table}[h]

\caption{The CcCR Vibration-Rotation Interaction Constants in MHz.}

\label{vib-rot}

\centering
\begin{tabular}{ c r r r }

\hline\hline
mode & \hspace{0.15in}$\alpha^A$ & \hspace{0.15in}$\alpha^B$ & \hspace{0.15in}$\alpha^C$ \\

\hline

1 &   6772.7 &  19.5 &   0.5 \\
2 &   6907.5 &  14.0 &   4.3 \\
3 &   -413.1 &  66.9 &  64.6 \\
4 &  -2487.6 &  -6.4 &  14.6 \\
5 &   1762.2 &  40.6 &   8.7 \\
6 & -10895.0 & -13.4 &  35.6 \\
7 & -28597.9 &  -7.9 & -23.2 \\
8 &  28365.9 & -47.6 & -16.8 \\
9 &  -7495.6 & -13.3 &  31.3 \\

\hline

\end{tabular}
\end{table}
\endgroup

\renewcommand{\baselinestretch}{1}
\begingroup
\begin{table}[h]

\caption{The CcCR QFF Quadratic and Cubic Force Constants (in
mdyn/\AA$^n$$\cdot$rad$^m$) for CH$_2$CN$^-$ in the symmetry-internal coordinate
system listed in Eqs.~2-10.}

\label{fc1}

\centering
\small

\begin{tabular}{c r c r c r c r c r}
\hline

$F_{11}$ &  5.530 104 & $F_{97}$ &  0.023 263 & $F_{521}$ &   0.2661 & $F_{651}$ &  -0.0943 & $F_{882}$ &  -0.1132 \\ 
$F_{21}$ &  0.128 241 & $F_{98}$ &  0.094 020 & $F_{522}$ &   0.0458 & $F_{652}$ &  -0.1221 & $F_{883}$ &  -0.1145 \\ 
$F_{22}$ &  6.555 883 & $F_{99}$ &  0.488 664 & $F_{531}$ &  -0.0496 & $F_{653}$ &  -0.0071 & $F_{884}$ &  -0.5701 \\ 
$F_{31}$ & -0.028 918 & $F_{111}$ & -22.6169 &  $F_{532}$ &   0.1555 & $F_{654}$ &  -0.4512 & $F_{885}$ &  -0.1370 \\ 
$F_{32}$ &  0.828 276 & $F_{211}$ &   0.1821 &  $F_{533}$ &  -0.0791 & $F_{655}$ &  -0.2330 & $F_{886}$ &  -0.0589 \\ 
$F_{33}$ & 15.588 505 & $F_{221}$ &  -0.4359 &  $F_{541}$ &   0.1369 & $F_{661}$ &  -0.0476 & $F_{971}$ &   0.0017 \\ 
$F_{41}$ &  0.230 077 & $F_{222}$ & -38.2328 &  $F_{542}$ &   0.1713 & $F_{662}$ &  -0.9049 & $F_{972}$ &  -0.1306 \\  
$F_{42}$ &  0.656 729 & $F_{311}$ &  -0.0215 &  $F_{543}$ &  -0.0559 & $F_{663}$ &  -0.7569 & $F_{973}$ &  -0.0740 \\  
$F_{43}$ & -0.167 307 & $F_{321}$ &   0.1660 &  $F_{544}$ &   0.4306 & $F_{664}$ &   0.0573 & $F_{974}$ &  -0.0367 \\  
$F_{44}$ &  0.429 388 & $F_{322}$ &  -2.0024 &  $F_{551}$ &  -0.1351 & $F_{665}$ &   0.0598 & $F_{975}$ &  -0.0010 \\  
$F_{51}$ &  0.259 999 & $F_{331}$ &  -0.1413 &  $F_{552}$ &   0.1132 & $F_{666}$ &  -0.1294 & $F_{976}$ &  -0.0421 \\  
$F_{52}$ &  0.140 187 & $F_{332}$ &  -0.6860 &  $F_{553}$ &  -0.0258 & $F_{771}$ & -22.7980 & $F_{981}$ &  -0.0507 \\  
$F_{53}$ & -0.132 967 & $F_{333}$ &-105.9486 &  $F_{554}$ &   0.1846 & $F_{772}$ &   0.2877 & $F_{982}$ &   0.0218 \\  
$F_{54}$ & -0.055 920 & $F_{411}$ &   0.2890 &  $F_{555}$ &  -0.5143 & $F_{773}$ &  -0.0386 & $F_{983}$ &  -0.1650 \\  
$F_{55}$ &  0.410 435 & $F_{421}$ &   0.0988 &  $F_{611}$ &   0.0250 & $F_{774}$ &   0.0134 & $F_{984}$ &  -0.1266 \\  
$F_{61}$ &  0.039 862 & $F_{422}$ &  -0.0823 &  $F_{621}$ &  -0.0069 & $F_{775}$ &   0.4085 & $F_{985}$ &  -0.0393 \\  
$F_{62}$ &  0.091 750 & $F_{431}$ &  -0.1456 &  $F_{622}$ &  -0.0968 & $F_{776}$ &   0.0044 & $F_{986}$ &   0.0626 \\  
$F_{63}$ &  0.039 511 & $F_{432}$ &   0.2752 &  $F_{631}$ &  -0.0291 & $F_{871}$ &   0.0772 & $F_{991}$ &  -0.0303 \\  
$F_{64}$ & -0.088 431 & $F_{433}$ &  -0.2106 &  $F_{632}$ &  -0.0076 & $F_{872}$ &  -0.3529 & $F_{992}$ &  -0.5724 \\  
$F_{65}$ & -0.098 023 & $F_{441}$ &  -0.0189 &  $F_{633}$ &  -0.0404 & $F_{873}$ &  -0.0494 & $F_{993}$ &  -0.8348 \\  
$F_{66}$ &  0.669 124 & $F_{442}$ &   0.0595 &  $F_{641}$ &  -0.1511 & $F_{874}$ &  -0.1370 & $F_{994}$ &  -0.0314 \\  
$F_{77}$ &  5.443 393 & $F_{443}$ &  -0.2530 &  $F_{642}$ &  -0.1286 & $F_{875}$ &  -0.0169 & $F_{995}$ &   0.0054 \\  
$F_{87}$ &  0.177 131 & $F_{444}$ &  -0.1887 &  $F_{643}$ &  -0.0403 & $F_{876}$ &  -0.0009 & $F_{996}$ &  -0.0664 \\  
$F_{88}$ &  0.528 308 & $F_{511}$ &   0.0395 &  $F_{644}$ &  -0.8662 & $F_{881}$ &  -0.1815 &           &          \\ 
                                                                       
\end{tabular}

\end{table}
\endgroup
\renewcommand{\baselinestretch}{2}

\renewcommand{\baselinestretch}{1}
\begingroup
\begin{table}[h]

\caption{The Quartic Force Constants (in mdyn/\AA$^n$$\cdot$rad$^m$) for
CH$_2$CN$^-$ in the symmetry-internal coordinate system for the CcCR QFF.}

\label{fc2}

\centering
\scriptsize

\begin{tabular}{c r c r c r c r c r}
\hline

$F_{1111}$ &  79.82 & $F_{5444}$ &   5.15 & $F_{6631}$ &  -0.49 & $F_{8755}$ &   0.07 & $F_{9811}$ &  -0.34  \\
$F_{2111}$ &  -0.90 & $F_{5511}$ &   0.40 & $F_{6632}$ &   1.20 & $F_{8761}$ &  -0.08 & $F_{9821}$ &   0.19  \\
$F_{2211}$ &  -1.61 & $F_{5521}$ &   0.09 & $F_{6633}$ &  -0.81 & $F_{8762}$ &   0.17 & $F_{9822}$ &  -0.62  \\
$F_{2221}$ &   0.79 & $F_{5522}$ &  -0.39 & $F_{6641}$ &   0.01 & $F_{8763}$ &   0.05 & $F_{9831}$ &  -0.22  \\
$F_{2222}$ & 181.08 & $F_{5531}$ &  -0.52 & $F_{6642}$ &  -0.05 & $F_{8764}$ &   0.11 & $F_{9832}$ &   0.17  \\
$F_{3111}$ &  -0.55 & $F_{5532}$ &   0.23 & $F_{6643}$ &  -0.23 & $F_{8765}$ &   0.14 & $F_{9833}$ &  -0.20  \\
$F_{3211}$ &   0.39 & $F_{5533}$ &  -0.13 & $F_{6644}$ &  -0.35 & $F_{8766}$ &  -0.39 & $F_{9841}$ &   0.03  \\
$F_{3221}$ &  -1.13 & $F_{5541}$ &   0.35 & $F_{6651}$ &   0.35 & $F_{8777}$ &  -0.87 & $F_{9842}$ &  -0.05  \\
$F_{3222}$ &   2.41 & $F_{5542}$ &   0.23 & $F_{6652}$ &  -0.28 & $F_{8811}$ &  -0.39 & $F_{9843}$ &   0.01  \\
$F_{3311}$ &  -0.55 & $F_{5543}$ &   0.02 & $F_{6653}$ &  -0.23 & $F_{8821}$ &   0.17 & $F_{9844}$ &  -0.58  \\
$F_{3321}$ &   0.31 & $F_{5544}$ &   3.63 & $F_{6654}$ &  -0.09 & $F_{8822}$ &  -0.72 & $F_{9851}$ &   0.13  \\
$F_{3322}$ &   2.93 & $F_{5551}$ &   0.94 & $F_{6655}$ &  -0.07 & $F_{8831}$ &  -0.20 & $F_{9852}$ &  -0.24  \\
$F_{3331}$ &  -0.27 & $F_{5552}$ &  -0.37 & $F_{6661}$ &  -0.10 & $F_{8832}$ &   0.22 & $F_{9853}$ &  -0.31  \\
$F_{3332}$ &  -5.04 & $F_{5553}$ &  -0.02 & $F_{6662}$ &   0.33 & $F_{8833}$ &  -0.25 & $F_{9854}$ &  -0.25  \\
$F_{3333}$ & 584.63 & $F_{5554}$ &   2.44 & $F_{6663}$ &   0.18 & $F_{8841}$ &   0.22 & $F_{9855}$ &  -0.09  \\
$F_{4111}$ &  -0.05 & $F_{5555}$ &   1.38 & $F_{6664}$ &  -0.07 & $F_{8842}$ &  -0.31 & $F_{9861}$ &  -0.14  \\
$F_{4211}$ &   0.31 & $F_{6111}$ &  -0.15 & $F_{6665}$ &  -0.07 & $F_{8843}$ &   0.31 & $F_{9862}$ &  -0.02  \\
$F_{4221}$ &  -0.10 & $F_{6211}$ &   0.13 & $F_{6666}$ &   2.94 & $F_{8844}$ &  -0.40 & $F_{9863}$ &   0.03  \\
$F_{4222}$ &   1.27 & $F_{6221}$ &  -0.06 & $F_{7711}$ &  82.01 & $F_{8851}$ &   0.22 & $F_{9864}$ &   0.14  \\
$F_{4311}$ &  -0.35 & $F_{6222}$ &  -0.04 & $F_{7721}$ &  -0.45 & $F_{8852}$ &  -0.41 & $F_{9865}$ &   0.11  \\
$F_{4321}$ &   0.12 & $F_{6311}$ &   0.00 & $F_{7722}$ &  -1.16 & $F_{8853}$ &   0.06 & $F_{9866}$ &  -0.18  \\
$F_{4322}$ &   0.03 & $F_{6321}$ &   0.04 & $F_{7731}$ &  -0.37 & $F_{8854}$ &   0.03 & $F_{9877}$ &  -0.24  \\
$F_{4331}$ &  -0.05 & $F_{6322}$ &   0.16 & $F_{7732}$ &   0.53 & $F_{8855}$ &   0.22 & $F_{9887}$ &  -0.30  \\
$F_{4332}$ &  -0.83 & $F_{6331}$ &  -0.18 & $F_{7733}$ &  -0.31 & $F_{8861}$ &  -0.07 & $F_{9888}$ &  -0.23  \\
$F_{4333}$ &   0.76 & $F_{6332}$ &   0.04 & $F_{7741}$ &  -0.18 & $F_{8862}$ &   0.05 & $F_{9911}$ &  -0.40  \\
$F_{4411}$ &  -0.28 & $F_{6333}$ &  -0.18 & $F_{7742}$ &   0.22 & $F_{8863}$ &   0.07 & $F_{9921}$ &   0.13  \\
$F_{4421}$ &   0.52 & $F_{6411}$ &  -0.05 & $F_{7743}$ &  -0.24 & $F_{8864}$ &  -0.01 & $F_{9922}$ &   0.07  \\
$F_{4422}$ &  -0.14 & $F_{6421}$ &   0.13 & $F_{7744}$ &   0.04 & $F_{8865}$ &   0.06 & $F_{9931}$ &  -0.24  \\
$F_{4431}$ &  -0.59 & $F_{6422}$ &   0.31 & $F_{7751}$ &  -0.12 & $F_{8866}$ &  -0.39 & $F_{9932}$ &   1.02  \\
$F_{4432}$ &   0.35 & $F_{6431}$ &   0.09 & $F_{7752}$ &  -0.01 & $F_{8877}$ &  -0.26 & $F_{9933}$ &  -0.66  \\
$F_{4433}$ &  -0.58 & $F_{6432}$ &  -0.13 & $F_{7753}$ &  -0.13 & $F_{8887}$ &  -0.16 & $F_{9941}$ &   0.01  \\
$F_{4441}$ &   0.65 & $F_{6433}$ &   0.22 & $F_{7754}$ &   0.28 & $F_{8888}$ &  -0.17 & $F_{9942}$ &   0.42  \\
$F_{4442}$ &   0.50 & $F_{6441}$ &  -0.80 & $F_{7755}$ &  -0.39 & $F_{9711}$ &  -0.43 & $F_{9943}$ &  -0.13  \\
$F_{4443}$ &   0.29 & $F_{6442}$ &  -1.41 & $F_{7761}$ &  -0.11 & $F_{9721}$ &  -0.02 & $F_{9944}$ &  -0.00  \\
$F_{4444}$ &   8.13 & $F_{6443}$ &   0.18 & $F_{7762}$ &   0.01 & $F_{9722}$ &  -0.06 & $F_{9951}$ &   0.28  \\
$F_{5111}$ &   0.04 & $F_{6444}$ &  -8.66 & $F_{7763}$ &   0.01 & $F_{9731}$ &  -0.29 & $F_{9952}$ &   0.03  \\
$F_{5211}$ &  -0.03 & $F_{6511}$ &  -0.03 & $F_{7764}$ &   0.06 & $F_{9732}$ &   0.29 & $F_{9953}$ &  -0.08  \\
$F_{5221}$ &   0.24 & $F_{6521}$ &   0.10 & $F_{7765}$ &   0.08 & $F_{9733}$ &  -0.23 & $F_{9954}$ &   0.05  \\
$F_{5222}$ &   0.29 & $F_{6522}$ &   0.19 & $F_{7766}$ &  -0.27 & $F_{9741}$ &   0.10 & $F_{9955}$ &   0.01  \\
$F_{5311}$ &  -0.66 & $F_{6531}$ &   0.09 & $F_{7777}$ &  83.38 & $F_{9742}$ &   0.12 & $F_{9961}$ &  -0.12  \\
$F_{5321}$ &  -0.01 & $F_{6532}$ &  -0.07 & $F_{8711}$ &  -0.79 & $F_{9743}$ &   0.03 & $F_{9962}$ &   0.09  \\
$F_{5322}$ &  -0.06 & $F_{6533}$ &   0.33 & $F_{8721}$ &   0.02 & $F_{9744}$ &   0.01 & $F_{9963}$ &   0.02  \\
$F_{5331}$ &   0.36 & $F_{6541}$ &  -0.54 & $F_{8722}$ &  -0.56 & $F_{9751}$ &   0.16 & $F_{9964}$ &  -0.11  \\
$F_{5332}$ &  -0.72 & $F_{6542}$ &  -0.90 & $F_{8731}$ &  -0.29 & $F_{9752}$ &  -0.10 & $F_{9965}$ &  -0.08  \\
$F_{5333}$ &   0.27 & $F_{6543}$ &   0.02 & $F_{8732}$ &   0.09 & $F_{9753}$ &  -0.22 & $F_{9966}$ &   0.64  \\
$F_{5411}$ &   0.20 & $F_{6544}$ &  -5.62 & $F_{8733}$ &  -0.33 & $F_{9754}$ &   0.15 & $F_{9977}$ &  -0.23  \\
$F_{5421}$ &   0.14 & $F_{6551}$ &  -0.62 & $F_{8741}$ &   0.13 & $F_{9755}$ &   0.07 & $F_{9987}$ &  -0.35  \\
$F_{5422}$ &  -0.04 & $F_{6552}$ &  -0.55 & $F_{8742}$ &   0.31 & $F_{9761}$ &  -0.16 & $F_{9988}$ &  -0.05  \\
$F_{5431}$ &  -0.17 & $F_{6553}$ &  -0.08 & $F_{8743}$ &   0.03 & $F_{9762}$ &   0.07 & $F_{9997}$ &  -0.30  \\
$F_{5432}$ &   0.07 & $F_{6554}$ &  -3.55 & $F_{8744}$ &  -0.05 & $F_{9763}$ &  -0.01 & $F_{9998}$ &  -0.36  \\
$F_{5433}$ &  -0.16 & $F_{6555}$ &  -2.02 & $F_{8751}$ &   0.31 & $F_{9764}$ &  -0.07 & $F_{9999}$ &   1.79  \\
$F_{5441}$ &   0.70 & $F_{6611}$ &  -0.53 & $F_{8752}$ &  -0.03 & $F_{9765}$ &  -0.01 & \\
$F_{5442}$ &   0.18 & $F_{6621}$ &   0.09 & $F_{8753}$ &  -0.33 & $F_{9766}$ &  -0.25 & \\
$F_{5443}$ &  -0.30 & $F_{6622}$ &   0.34 & $F_{8754}$ &   0.06 & $F_{9777}$ &  -0.45 & \\

\end{tabular}

\end{table}
\endgroup
\renewcommand{\baselinestretch}{2}

\end{document}